\documentclass{article}
\usepackage{amsmath,amsfonts}

\usepackage{graphicx}
\usepackage{textcomp}
\usepackage{xcolor}
\usepackage{algorithm}
\usepackage{algpseudocode}
\usepackage{pifont}
\usepackage{amsthm}
\usepackage{comment}
\usepackage{amssymb}
\usepackage{soul}

\def\BibTeX{{\rm B\kern-.05em{\sc i\kern-.025em b}\kern-.08em
    T\kern-.1667em\lower.7ex\hbox{E}\kern-.125emX}}
    
    \usepackage{tikz}

\usepackage{listings}

\usepackage{tikz}
\usetikzlibrary{arrows,fit,shapes}
\usetikzlibrary{calc}
\tikzset{fontscale/.style = {font=\relsize{#1}}}
\usepackage{pgfplots}
\usepackage{url}
\newcommand{\deadline}{\mathsf{D}}
\newcommand{\period}{\mathsf{T}}

\ifdefined\comments
\newcommand{\is}[1]{\textcolor{blue}{[Nacho: #1]}}
\newcommand{\nc}[1]{\textcolor{magenta}{[Nicola: #1]}}
\newcommand{\he}[1]{\textcolor{violet}{[Houssam : #1]}}
\newcommand{\js}[1]{\textcolor{red}{[Jayati: #1]}}
\newcommand{\sef}[1]{\textcolor{orange}{[Sebastien: #1]}}
\newcommand{\deleted}[1]{\st{#1}}
\else 
\newcommand{\is}[1]{\textcolor{blue}{[]}}
\newcommand{\nc}[1]{\textcolor{magenta}{[]}}
\newcommand{\he}[1]{\textcolor{violet}{[]}}
\newcommand{\js}[1]{\textcolor{red}{[]}}
\newcommand{\sef}[1]{\textcolor{orange}{}}
\newcommand{\deleted}[1]{}
\fi

\ifdefined\shepherding
\newcommand{\newtext}[1]{{\color{red} #1}}
\else
\newcommand{\newtext}[1]{#1}
\fi

\newtheorem{lemma}{Lemma}
\newtheorem{definition}{Definition}

\begin{document}

\title{Contention-Aware GPU Partitioning and Task-to-Partition Allocation for Real-Time Workloads}

\author{Houssam-Eddine Zahaf, Ignacio Sañudo Olmedo, Jayati Singh, \\ Nicola Capodieci, Sébastien Faucou}

\maketitle

\begin{abstract}
In order to satisfy timing constraints, modern real-time applications require massively parallel accelerators such as General Purpose Graphic Processing Units (GPGPUs). Generation after generation, the number of computing clusters made available in novel GPU architectures is steadily increasing, hence, investigating suitable scheduling approaches is now mandatory. Such scheduling approaches are related to mapping different and concurrent compute kernels within the GPU computing clusters, hence grouping GPU computing clusters into schedulable partitions. In this paper we propose novel techniques to define GPU partitions; this allows us to define suitable task-to-partition allocation mechanisms in which tasks are GPU compute kernels featuring different timing requirements.
Such mechanisms will take into account the interference that GPU kernels experience when running in overlapping time windows. Hence, an effective and simple way to quantify the magnitude of such interference is also presented.
We demonstrate the efficiency of the proposed approaches against the classical techniques that considered the GPU as a single, non-partitionable resource. 
\end{abstract}

{\bf keywords} {GPU,
GPU partitioning,
task allocation,
Real-time,
Memory Interference.
}

%

\section{Introduction}
The increasing complexity of automotive architectures is posing
new challenges for the integration of these technologies in safety
related domains. Following the recent advancements in the ADAS (Advanced Driving Assistance System) industry, many tier-1 \newtext{suppliers} and OEMs \newtext{(Original Equipment Manufacturer)} are moving towards deploying their applications within heterogeneous platforms.
Such architectures are usually based on multi-core CPUs coupled with massively parallel accelerators.  
While the adoption of such heterogeneous platforms 
satisfies the computational needs for ADAS-related workloads, 
engineering difficulties arise when the system is subject to real-time and safety requirements \cite{burgio2017software, olmedo2018perspective,challenge}.

In this context, Graphic Processing Units (GPUs) released by NVIDIA have been recently deployed in many latency-sensitive domains. The NVIDIA CUDA API allows the programmer to easily and efficiently exploit the processing power enabled by these massively parallel and general purpose accelerators.
The CUDA execution model enables a CPU Host to offload \textit{compute kernels} to the GPU. 
\newtext{A compute kernel is a parallel task.}
The programmer can elect to submit such kernels in a FIFO fashion, or through concurrent execution, in which kernels are offloaded through \textit{parallel streams}.
Experimental results on both early and recent NVIDIA architectures have shown counter-intuitive results: when executing several independent kernels, parallel flows might result in sub-optimal predictability and latency compared to the baseline FIFO execution \cite{sched_hierarchy_RTAS}. This is due to the fact that independent CUDA streams offloaded to a GPU will contend memory and compute resources. 
Moreover, the programmer typically has very little control on how to map kernels/streams to the GPU compute resources~\cite{otterness2020amd}. 

In NVIDIA GPUs, compute resources are organized in \emph{Streaming
Multiprocessors} (SMs); a SM is a cluster of Arithmetical Logical Units (CUDA cores). The number of SMs is steadily increasing generation after generation. 
The Fermi and Kepler GPU architectures, released in 2010 - 2012, 
featured around 15 SMs where each SM featured 32 CUDA
cores in Fermi and 192 in Kepler. The 2014 Maxwell architecture featured up to 24 SMs with 128 cores per SM, while the subsequent generations
Pascal, Volta and Turing feature around 50 and 80 SMs
which in turn hold 64 and 128 CUDA cores per SM respectively. The most recently released architecture (Ampere) integrates more than 100 SMs with 64 CUDA cores per SM.

The typical usage of the GPU as a single resource leads to an under-utilization of its computing power. This has been shown in many works in the Neural Network literature \cite{jain2018dynamic}  
and in different well-known benchmarks \cite{elastic} such as ${\sf Parboil}$ \cite{stratton2012parboil} or ${\sf Rodinia}$ \cite{che2009rodinia}. 
Resource under-utilization and unpredictability are further exacerbated as the number of available computing clusters in a GPU increases. 
It becomes critical for the real-time system engineer to correctly define and enforce partitions of the GPU SMs among kernels characterized by different timing requirements. 

\newtext{To mitigate the under-utilization problem, concurrent multi-kernel execution is a viable option. This can be achieved by allowing multiple kernels to reside within the same SMs in overlapping time windows. This is known as Simultaneous Kernels (\textit{SMK}), and it is only possible when the involved kernels satisfy specific occupancy conditions \cite{sched_hierarchy_RTAS}. A different approach, that is usually referred as \textit{spatial multikernel} \cite{gpu_multitasking}, allows the programmer to map different kernels to different groups or partitions of SMs.} 


\newtext{The standard CUDA execution model does not allow to spatially partition the computing resources nor the execution of concurrent kernels from different processes.
For this purpose, NVIDIA provides the CUDA Multi-Process Service (MPS). CUDA MPS is a software feature that allows multiple processes in different contexts to execute concurrently and to reserve a percentage of the GPU computing resources to specific applications (clients).}

However, this feature presents many drawbacks in terms of spatial isolation. For instance, the memory hierarchy (i.e. memory bandwidth and caches) remains shared and therefore contended among all the MPS clients. With the advent of the Ampere architecture, a new feature called Multi-Instance GPU (MIG) was announced.
MIG allows the system engineer to effectively partition
the memory and compute resources of the GPU. \newtext{Specifically, the partitions can be defined at the granularity of a GPC (Graphic Processing Cluster, which groups multiple SMs). For example, the A100 GPU features 7 GPCs so that it is possible to configure the GPU with 7 static partitions each using 1 GPC, each partition having a dedicated use of the compute resources (SMs) and memory resources (L2/LLC slices). More details about the possible MIG configurations are publicly available in the relevant NVIDIA documentation}\footnote{\url{https://docs.nvidia.com/datacenter/tesla/mig-user-guide/index.html}}.

In literature, currently available GPU partitioning mechanisms are designed to provide a coarse grained quality of service, rather than providing real-time guarantees. \newtext{The previously mentioned MPS and MIG enable different levels of GPU partitioning, however, scheduling aspects are still a responsibility of the system engineer.}
In this paper we \newtext{therefore} propose novel GPU partitioning heuristics based on the allocation of GPU partitions according to the real-time constraints of the tasks. This implies accounting for the interference that the different tasks might experience when using shared resources. Our proposed mechanism (i) defines every partition size, (ii) allocates the real-time tasks to every partition so that the timing constraints of the tasks are met. 

This paper is organized as follows: in
Section \ref{programmingModel}, we report a description of the CUDA programming model. 
In Section \ref{kernelProperties} we present the techniques to estimate and model the interference among GPU kernels. 
In Section \ref{systemModel} we formalize the model of the architecture and the workload.
Section \ref{heuristics} presents the different heuristics to assign kernels to GPU partitions. In section \ref{relatedWork} we present the related work in this domain. Results and conclusive discussions are reported in sections \ref{results} and \ref{conc}.

\section{CUDA programming model}\label{programmingModel}
\begin{figure}
\centering
  \includegraphics[width=\linewidth]{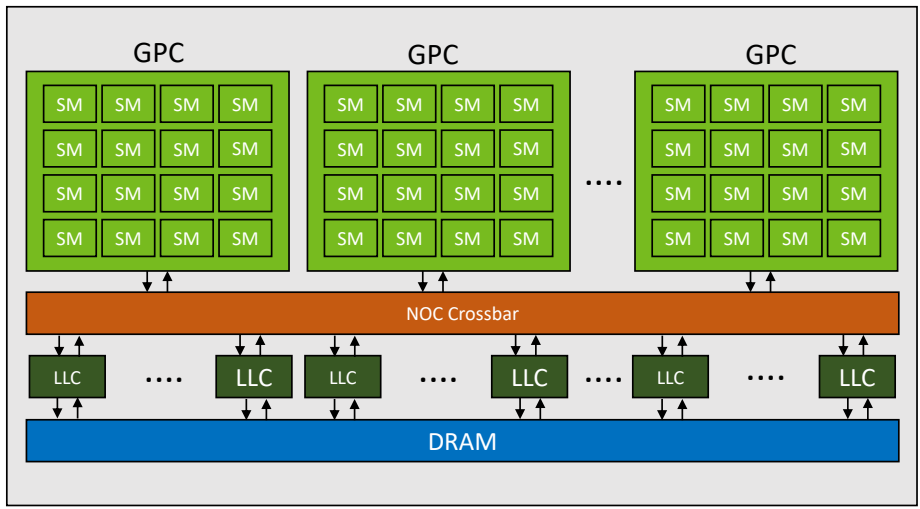}
  \caption{GPU architecture design.}
  \label{fig:gpu_arch}
\end{figure}

A NVIDIA GPU integrates several Streaming Multiprocessors (SMs) and
one or more copy engines (CEs). \newtext{Streaming multiprocessors are
  a collection of compute resources having the ability to execute CUDA
  kernels. Streaming multiprocessors are grouped into Graphic
  Processing Clusters (GPC). For instance, as depicted in Figure
  \ref{fig:gpu_arch}, each GPC is composed of 16 SMs (as in the new Ampere
  NVIDIA GPU, the A100) with a dedicated memory controller and
  interface towards the memory hierarchy}. Whereas copy engines
perform memory copy operations between different address spaces. A GPU
can be programmed using GPGPU APIs such as OpenCL, which is an open
standard, or the NVIDIA CUDA proprietary standard. From the point of
view of the programmer, these APIs expose a similar set of
functionalities.  A typical programming pattern for GPGPU programs
works as follows.  First, memory allocation operations are performed
both on CPU (host) and GPU (device) side. In a real-time setting, such
allocations are performed as initialization
procedures. 
Then, memory copies are operated between the main memory and the GPU
accessible memory. One ore more kernels are then launched and executed
on the GPU, and the results are eventually copied back to the main
host memory.

Programming a GPU kernel \newtext{(also named CUDA or compute kernel)}
requires dividing the parallel algorithm into a compute grid. A grid
is itself composed of several blocks. A thread block is a set of
threads that can be executed serially or in parallel. Multiple blocks
can be allocated in a single SM. The number of thread blocks per grid
and the number of threads per block is defined by the programmer and
they can be organized in 1D, 2D and 3D dimensions. The thread block
scheduler distributes the blocks to the different SMs in a round-robin
fashion. However, the programmer can elect to use CUDA streams to
dispatch multiple kernels. A CUDA stream is an abstraction of a
sequence of compute and copy operations: operations within the same
stream are executed in the order they are submitted, whereas commands
of different streams are independent and thus operations belonging to
different streams might overlap in time. If streams are considered,
the GPU block scheduler dispatches thread blocks from the different
kernels among the SMs considering the occupancy of memory and compute
resources requested by the submitted
kernels~\cite{sched_hierarchy_RTAS}.  In NVIDIA architectures, blocks
are composed of warps which in turn are composed of 32 threads. A warp
is the minimum schedulable unit \newtext{and all the threads within
  a warp execute the same instruction}. Each SM is composed of
multiple warp schedulers (e.g. up to 4 in the last generation
architectures). The warp scheduler is in charge of distributing the
threads to the available hardware resources organizing
ready-to-execute memory or compute instructions from the different
available and ready warps.
\section{GPU Kernels properties}\label{kernelProperties}

Memory access patterns of CUDA kernels can drastically impact timing performance. According to its characteristics and its functional behavior, a CUDA kernel might feature different compute and memory resource requirements \cite{elastic}. 
According to such requirements, kernels can be classified into two categories: \emph{compute intensive} and \emph{memory intensive}.
Intuitively, a kernel is said to be memory intensive if a consistent chunk of its execution time is spent in accessing memory. Conversely, it is defined as compute intensive, when most of its cycles are spent in computing instructions.

Co-running kernels compete on shared resources, hence, their executions interfere one with each other, hindering their overall performance.
More specifically, if kernels are running in parallel, the magnitude of performance deterioration caused by mutual interference depends on the category in which such kernels are classified. 
For instance, if two parallel kernels are classified as memory intensive, the overall system will experience strong contention on the memory hierarchy, causing notable performance deterioration. Such a decrease in performance would be significantly lower in the case the two kernels were of different types.

It is important to find an accurate model based on which kernels can be categorized as compute or memory intensive, so that contention-aware scheduling policies can be implemented.

\subsection{Kernel characterization}

Classifying a kernel requires the characterization of its memory access pattern, which can be achieved by analyzing CUDA- or binary-kernel code. However, it requires a tedious effort and a strong knowledge on GPU architectures and internals. In this paper, we use a simple, yet efficient approach to determine whether a kernel is compute or memory intensive.



It is known \cite{volkov, mei2016dissecting} that load and store instructions to global memory (i.e. LLC and main memory) present a significantly higher latency compared to compute instructions. Hence, as a rule of thumb, kernels characterized by large working data sets with few compute instructions will fall into the memory intensive category. Also, the performance of kernels belonging to such a category will therefore more strongly depend on the actual memory bandwidth allocated to them, rather than the compute resources (i.e. SMs) assigned to them.  

\begin{figure}[ht]
  \centering
  \resizebox{0.99\columnwidth}{!}{
    \begin{tikzpicture}[every mark/.append style={mark size=1pt}]
      \begin{axis}
        [
        axis x line = bottom,
        axis y line = left,
        xlabel={Number of used SMs},
        ylabel={Execution time [ms]},
        xmax = 50,
        ymax = 1.5,
        xmin = 0,
        ymin = -0.1,
        y label style={at={(axis description cs:-0.07,.5)},rotate=0,anchor=south},
        legend entries={HS,PF, TRNS, VADD},
        legend style={at={(0.8,1)}} ]
        \addplot table[y index=1,x index=0]{texfiles/results/reswcet.dat};
        \addplot table[y index=2,x index=0]{texfiles/results/reswcet.dat};
        \addplot table[y index=3,x index=0]{texfiles/results/reswcet.dat};
        \addplot table[y index=4,x index=0]{texfiles/results/reswcet.dat};

      \end{axis}
    \end{tikzpicture} 
    }
  \caption{Normalized Execution time of Hotspot (HS), Pathfinder, Matrix Transpose (TRNS) and Vector Add (VADD) as a function of the number of SMs}\label{fig:wcet}
 
\end{figure}
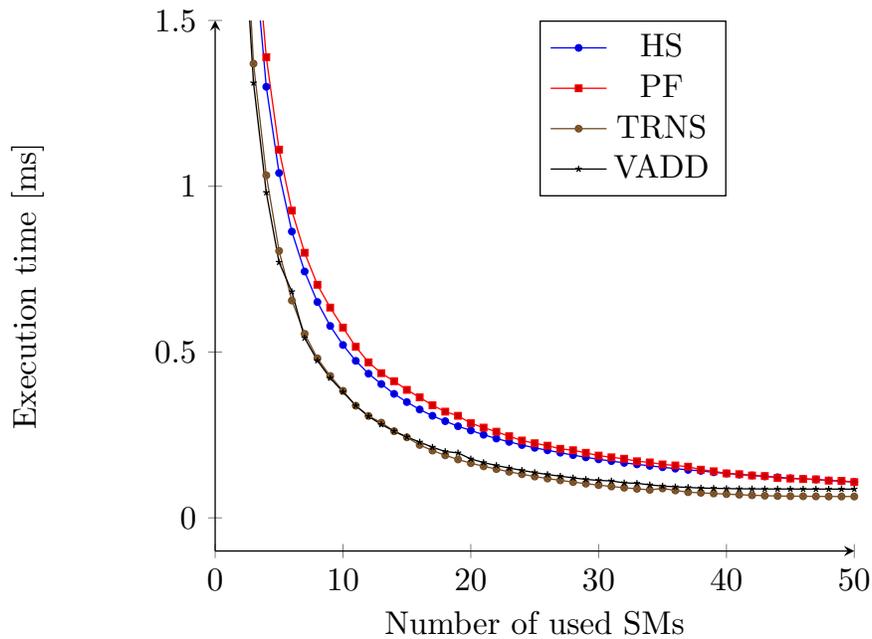

In order to further understand how performance in memory and compute kernels is affected by the number of allocated SMs and memory bandwidth,
we execute four benchmark kernels varying the number of SMs assigned to each. More specifically, Figure 
\ref{fig:wcet} reports the execution times of CUDA kernels extracted from the CUDA API Samples and Rodinia \cite{che2009rodinia}, these are: ${\sf vector\_add}$ (VADD), ${\sf matrix\_transpose}$ (TRNS), ${\sf hotspot}$ (HS) and  ${\sf path\_finder}$ (PF).
The execution time of each kernel has been normalized with respect the execution time of that same kernel running on a single SM.


Experiments in Figure \ref{fig:wcet} have been measured on a NVIDIA GeForce RTX2080 Ti GPU, featuring 68 SMs. 
In this experiment we first launch a dummy kernel on a CUDA stream.
The dummy kernel occupies a known and configurable number of SMs, preventing other CUDA streams (and related kernels) to occupy the same SMs as the dummy kernel. 
For instance, if we want to allocate $m$ number of SMs to the kernel under test, we launch the dummy kernel so as to occupy $68-m$ resources. In our experiments, we then launch the desired kernel on the unused $m$ SMs by simply exploiting a different CUDA stream.

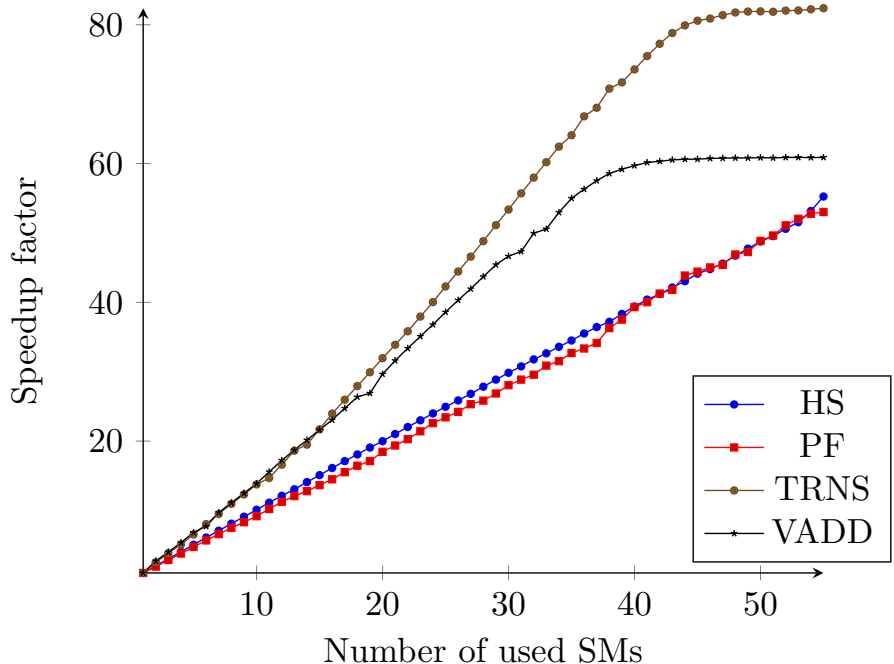
\begin{figure}[ht]
  \centering
  \resizebox{0.99\columnwidth}{!}{

    \begin{tikzpicture}[every mark/.append style={mark size=1pt}]
      \begin{axis}
        [
        axis x line = bottom,
        axis y line = left,
        xlabel={Number of used SMs},
        ylabel={Speedup factor},
        xmax = 55,
        y label style={at={(axis description cs:0.05,.5)},rotate=0,anchor=south},
        legend entries={HS,PF, TRNS, VADD},
        legend style={at={(1.1,0.35)}} ]
        
        \addplot table[y index=1,x index=0]{texfiles/results/speedupfactors.dat};

        \addplot table[y index=2,x index=0]{texfiles/results/speedupfactors.dat};
        
        \addplot table[y index=3,x index=0]{texfiles/results/speedupfactors.dat};
             
        \addplot table[y index=4,x index=0]{texfiles/results/speedupfactors.dat};

      \end{axis}
    \end{tikzpicture}}
  \caption{speed up factors  of Hotspot (HS), Pathfinder, Matrix Transpose (TRNS) and Vector Add (VADD) as a function of the number of SMs}\label{fig:speed}
\end{figure}

\noindent Even if the execution time of each kernel in Figure \ref{fig:wcet} has been scaled to have the same one SM-execution time, their execution time varies from kernel to kernel as we increase the number of SMs. Nevertheless, we highlight two categories of behavior. The first includes the ${\sf vec\_add}$ and the ${\sf transpose}$ kernels, which are known to be memory intensive, whereas kernels ${\sf hotspot}$ and ${\sf pathfinder}$, can be considered as compute intensive. These results are also confirmed by Figure \ref{fig:speed}.

\noindent In Figure \ref{fig:speed}, we report the speedup factors of the different kernels as a function of the number of assigned SMs. When the number of SMs is less than 40, the increase of the number of SMs leads to linearly reduce the kernels' execution times. Starting from 40 SMs, the execution times of memory intensive kernels do not increase anymore, while those of compute intensive kernels keep improving linearly. In fact, for the memory intensive kernels, the GPU memory transfer bandwidth reaches saturation, using only 50\% of the compute resources (SMs). As our experiments show, this does not occur in compute intensive kernels. These considerations lead to the following definition.

\begin{definition}
Given an architecture, a kernel is considered to be memory intensive if, beyond a certain threshold, allocating more SMs does not improve its performance on that architecture.
Conversely, a kernel which is not memory intensive is considered to be compute intensive.
\end{definition}

\newtext{We can further improve the accuracy of the definition by considering other profiling metrics such as cache hit/miss rate, however this remains as a future work.}

\subsection{Performance and interference among co-running kernels}

Given real-time constraints, we aim to assign different GPU partitions to different co-running kernels. With GPU partition we indicate the set of SMs in which we allow kernels to run.
Contention-aware GPU partition allocation requires to (i) quantify the interference experienced by co-running kernels, and (ii) the definition of the different parameters that may impact the performance. 




\begin{figure}
  \centering
  \resizebox{\columnwidth}{!}{
  \begin{tikzpicture}
    \node at (0,0){\includegraphics{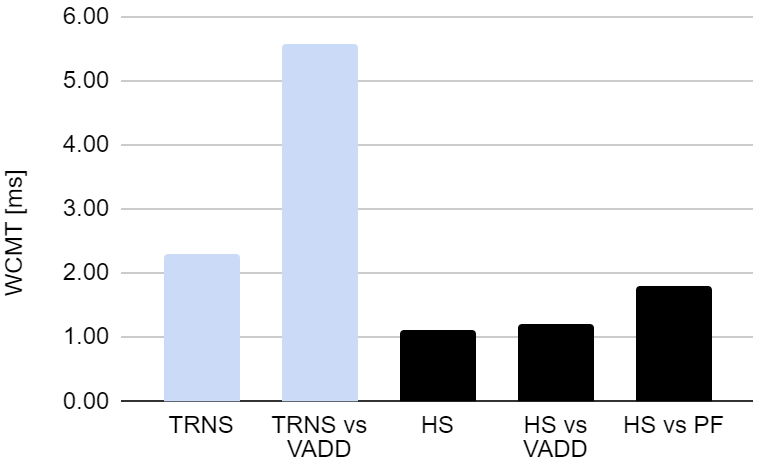}};
    \node at (-2.4,0) {\bf 2.229};
    \node at (-0.85,2.7) {\bf 5.548};
    \node at (0.7,-1) {\bf 1.091};
    \node at (2.25,-0.95) {\bf 1.184};
    \node at (3.8,-0.36) {\bf 1.77};
  \end{tikzpicture}
  }
  \caption{TRNS and HS WCMTs (Worst Case Measured Time) with and without interference from VADD and PF kernels.}
  \label{fig:interfhisto}
\end{figure}
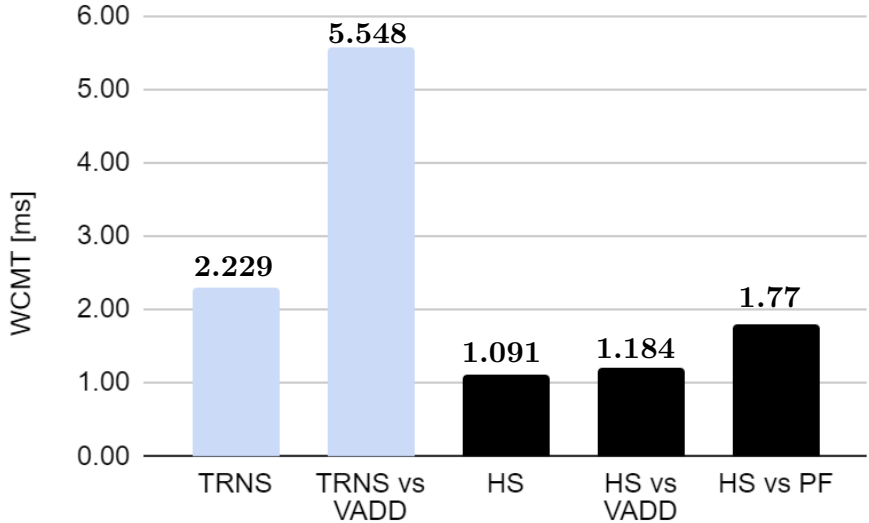

We aim to characterize and define the magnitude of interference that a kernel experience according to the tasks that are allocated with it on the same partition. 
In Figure \ref{fig:interfhisto}, our experiments are set so that each kernel occupies exactly 50\% of GPU resources. Specifically, this means that each kernel occupies half of all the SMs. This is possible because, depending on the kernel launch configuration (i.e., grid/blocks/threads configuration) more than one block can be scheduled concurrently within the same SM.
The first analysed kernel is the matrix transpose (${\sf TRNS }$) in isolation and interfered with the vector add. Its worst case measured execution time is equal to $2.300$ ms, and the average value is equal to $2.121$ ms. 
However, when the same kernel - which is memory intensive - runs with another memory intensive kernel (${\sf VADD }$), its execution time dramatically increases, as it goes from $2.300$ ms to $5.566$ ms. This decrease in performance quantifies to more than twice its execution time in isolation. Similarly, we run the hotspot compute kernel (${\sf HS}$), with and without the interference coming from other compute and memory intensive kernels (${\sf VADD }$ and pathfinder, ${\sf PF }$). Same observations can be drawn for such kernels, however, the performance drop is less dramatic. Nevertheless, performances do not significantly downgrade when a compute intensive kernel is executed along with a memory intensive kernel. Indeed, it has been shown in \cite{elastic}, that compute and memory intensive kernels interleave with minor effects on their execution times when they are executed concurrently. Therefore, it is important to define a proper strategy to allocate tasks in the same or different partitions, to reduce or eliminate the interference. For example, allocating kernels of different types to the same partition(s), thus to reduce contention on similar resources can help to increase the platform's utilization and predictability.

\subsection{Modeling kernel execution times}
\newtext{
The measurements presented above give an idea of the timing behaviour of compute kernels running on a GPU.
In the following, we present a heuristic for partitioning GPU compute resources, and evaluate it on a synthetic dataset.
To produce this dataset, we adopt a very simple model to describe these behaviours.
This model does not aim to describe in detail the various phenomena at work, but simply to produce inputs similar to those observed in the measurements, which is sufficient to evaluate the heuristic satisfactorily.}

\newtext{First, we observe that the kernel execution time scales as a semi-linear function of the total number of SMs.
Next, we observe that, for a constant number of allocated resources, the execution time of a kernel is more impacted when it is in the presence of a kernel that uses mostly the same type of resource. Rather than explicitly modeling the interference, we use two different functions for each kernel, corresponding for the first one to an execution with no or little interference, and the second one to an execution with a lot of interference.}

\[ 
\mathcal{C}^n(m) = \frac{a^n}{m} + b^n
\]

\[
\mathcal{C}^c(m) = \frac{a^c}{m} + b^c
\]

\noindent \newtext{Where $m$ is the number of SMs, $a^n$ and $b^n$ (resp. $a^c$ and $b^c$) are constants describing the execution of the kernel with no or little interference (resp. with a lot of interference). 
These functions will be used in the experimental section to evaluate the performances of our heuristics.
}

\section{System Model}\label{systemModel}
The goal of this paper is to define how a GPU can be partitioned, and how kernels can be allocated to the different partitions. The objective is to reduce interference among the different kernels, especially those of the same type.

\subsection{Architecture Model}

A GPU is composed of ${\sf M}$ streaming multiprocessors (SMs). The $j^{\text{th}}$ SM is denoted by $p_j$. A partition is a set of SMs. The $k^{\text{th}}$ Partition $\mathcal{P}_k = \{p_{k,1}, p_{k,2}, \cdots, p_{k,{|\mathcal{P}|}}\}$ is a composed of $|\mathcal{P}_k|$ SMs. Partitions do not share compute resources (i.e, $\mathcal{P}_k \cap \mathcal{P}_{k'} = \varnothing, \forall k, k \neq k'$). 

\newtext{We make the hypothesis that the considered architecture does not lead to timing anomalies. In particular, we hypothesize that the response time of a job cannot decrease when more SMs are assigned to it. This hypothesis is consistent with the measurements illustrated by Figure~\ref{fig:speed} which shows that once memory becomes the bottleneck, performance stagnates but does not decrease.}

\subsection{Task Model}

Let $\mathcal{T}=\{ \tau_1, \tau_2, \cdots, \tau_n \}$ be a set of kernels, henceforth \textit{tasks}. Each task $\tau_i$ is a sequence of jobs (possibly infinite),
characterized by a tuple
$\tau_i = \{\period_i, \deadline_i, \mathsf{M}_i,
\mathcal{C}_i^c(m), \mathcal{C}_i^{\sf n} (m) \}$, where:

\begin{itemize}
\item $\mathsf{T}_i$: is the task period. It represents the minimum time between two successive activations of task $\tau_i$.
\item $\mathsf{D}_i$ : is the relative deadline; Task $\tau_i$ must complete no later than $\mathsf{D}_i$ time units from its arrival.

\item $\mathcal{C}_i^c(m)$ : 
represents the task
  execution time when the task has at least one conflict and is
  running on all streaming multiprocessors on a partition of size $m$.
  A memory intensive (resp. compute intensive) task is said \textit{to be in conflict}, if at least another memory intensive (resp. compute intensive) kernel is allocated within the same partition.

\item $\mathcal{C}_i^{\sf n} (m)$ : represents the task execution time when it is not in the previous situation, and on partition of size  $m$.

\item $\mathsf{M}_i$ : Denotes the task type. If the task is memory intensive, $\mathcal{M}_i$ is set to ${\sf memory}$, otherwise it is set to ${\sf compute}$.
  
\end{itemize}

We denote by $\mathcal{T}(\mathcal{P})$ the set of tasks that are allocated to partition $\mathcal{P}$.
We denote by $\mathcal{C}_i(\mathcal{T}(\mathcal{P}), |\mathcal{P}|)$
the execution time of task $i$ when allocated with the set of tasks
$\mathcal{T}$, thus:

\[
  \mathcal{C}_i(\mathcal{T}(\mathcal{P}), |\mathcal{P}|)=\begin{cases}
                \mathcal{C}_i^c(|\mathcal{P}|), ~\text{if}~ \tau_i~\text{has}~{\sf conflict}  \\
                \mathcal{C}_i^{\sf n} (|\mathcal{P}|),~\text{otherwise}
            \end{cases}
\]

$\tau$ can have ${\sf conflict}$, according to its type, and the number of tasks allocated with it on the same partition, as well as their type.


For example, let us a consider a GPU partition in which we assign 10 kernels, 9 of which are compute intensive. The remaining memory intensive kernel will not experience any contention and its conflict \emph{tag} is set for non-conflict, whereas the nine compute kernels are in a state of conflict.


\section{Heuristics}\label{heuristics}

In the considered architecture, there are ${\sf M}$ compute resources (SMs),
and our goal is to allocate tasks (kernels) to compute resources such that the overall interference is reduced, and all deadlines are met. The addressed problem is quite challenging: it takes as
input a set of tasks, and it requires (i) partitioning the M resources
to a set of partitions, selecting the suitable partition size; (ii) allocating tasks to partitions, so that no deadline is
missed with the objective of reducing the overall interference. The space of solutions is highly combinatorial, as both the SM-to-partition and the task-to-partition
allocations are variable.

\newtext{A simpler problem of considering only the task-to-partition allocation has been shown to be intractable and NP hard for large numbers of resources and tasks.   In this paper, we tackle a more complex problem as it includes not only the task to partition allocation, but as well defining the partition size. }
Therefore, we investigate tractable heuristics aiming to explore the space of possible
solutions with a reasonable complexity (pseudo polynomial complexity of $\theta(M \cdot N \cdot H)$\footnote{M represents the number of SMs, N the total number of tasks $H$ task set hyperperiod}.
  We highlight that we may omit the task index when it is not necessary. 

\begin{algorithm}[h]
\caption{Partition and allocate}
\label{CHalgorithm}
\begin{algorithmic}[1]
  \State {\bf input : } Taskset $\mathcal{T}$, integer ${\sf M}$
  \State ${\sf par\_list = init\_partitions()}$
  \State {\bf if} (exhaustive) ${\sf fill\_forbidden\_list }$
  \While {($\Pi > {\sf M  } $)} 
       \State ${\sf (\mathcal{P}^1, {\sf elig\_list}) = select\_partitions(par\_list)}$
       \If {(${\sf \mathcal{P}^1== NULL}$)}
            \State {\bf return} ${\sf false}$ 
        \EndIf
        \Repeat
        \State ${\mathcal{P}^2 = {\sf choose}({\sf elig\_list})}$
        \State ${\sf elig\_list} = {\sf elig\_list} \setminus \{ \mathcal{P}^2 \}$
        \State ${\sf  \mathcal{P}^3 = merge(\mathcal{P}^1,\mathcal{P}^2) }$
        \If {$\sf (\mathcal{P}^3 == NULL)$}
        \State ${\sf add\_to\_forbidden\_moves({\sf \mathcal{P}^1},{\sf \mathcal{P}^2});}$
        \Else
            \State ${\sf par\_list = par\_list \setminus \sf \{\mathcal{P}^1, \mathcal{P}^2\}}$
            \State ${\sf par\_list = par\_list \cup \{ \mathcal{P}^3\}}$
            \EndIf
        \Until $(\mathcal{P}^3 \neq {\sf NULL}) \lor ({\sf elig\_list} = \varnothing)$
        \EndWhile
        \State {\bf return} ${\sf true}$
\end{algorithmic}
\end{algorithm}

Before starting our heuristic, we apply a fast necessary
schedulability test described in Lemma \ref{lemma:sched:necessary}

\begin{lemma}\label{lemma:sched:necessary}
  Let $\mathcal{T}$ be a task set.

  $\mathcal{T}$ is not schedulable if : 

  \[
     \sum_{i}^{|\mathcal{T}|} \frac{\mathcal{C}^n_i({\sf 1})}{{\sf T}_i} > M
  \]
  
\end{lemma}
\begin{proof}
The proof comes straight forward from classical schedulability tests based on utilization for multicore architectures. 
  

\end{proof}

The proposed heuristic is greedy, but keeps trace of different solutions that
have failed. It starts by applying Lemma \ref{lemma:sched:necessary} to
check if the task is not schedulable (not explicitly mentioned on the algorithm). The pseudo-code of the main
procedure is shown in Algorithm \ref{CHalgorithm}. The algorithm starts by 
building a baseline set of partitions (${\sf par\_list}$), where each
contains only one task, (Line 2). This list is maintained in a given
order (decreasing/increasing utilization order). At each step, the
algorithm selects two partitions from the partition list (Line 4) and
tries to merge them in order to reduce the number of required
resources (Line 11).  If the merge is not possible, the algorithm adds
a trace of the impossible allocation (Line 13), and try another couple
of partitions. The algorithm exits on ${\sf success}$ if the total
number of compute resources in ${\sf par\_list}$ is not greater than the number
of available resources, or it exits on ${\sf fail}$ if it can not select merge
candidates that would allow to decrease the number of required resources.

\subsection{Init partitions}

Our heuristic starts by building the baseline set of partitions ${\sf
  par\_list}$. Specifically, for every task it creates a partition. As the
task executes \newtext{without contending with other tasks}, the task execution time array
$\mathcal{C}_i^n(\mathcal{P})$ is considered. \newtext{However, threads of the same task compete on the resources within a SM. The contention overheads in this case are included within the task WCET.}

\begin{lemma}\label{lemma:tri}
  Let $\mathcal{P}$ be partition, such that $\mathcal{T}_{\mathcal{P}} = \{\tau\}$.
  
  Task $\tau$ is schedulable under $|P|$ resources if :

\[
  |\mathcal{P}| = \min \{m \in \{1, \cdots, {\sf M}\},  \text{such that } C^{n}(m) -   {\sf D} \leq 0, \},
\]
  
\end{lemma}

\begin{proof}
  The execution time is a decreasing function of the number of
  resources. Therefore, to ensure the task schedulability
  is sufficient to set its partition size to the smallest number of
  compute resources, allowing the task worst case execution time to be
  equal or less than the task deadline.  Therefore, the partition size
  is set to the first number of SMs that verifies the latter condition.
\end{proof}

The goal of this step is to create for every task a partition. The
condition is that the partition has the minimum number of SMs, such
that the task is schedulable. Lemma
\ref{lemma:tri} is applied on every task, and every produced partition
is stored in ${\sf par\_list}$.

At every step of our heuristic, we ensure that every partition within ${\sf par\_list}$ is schedulable.

We denote by $\Pi$ the number of compute resources that are required
in ${\sf par\_list}$. It is computed as follows:

\[
\Pi = \sum_{l=1}^{\sf |par\_list|} |\mathcal{P}_l|
\]

\begin{lemma}\label{lemma:sched:all}
  Let $\mathcal{T}$ be the task set partitioned using Algorithm
  \ref{CHalgorithm}.

  $\mathcal{T}$ is schedulable, if :
\[
  \Pi \leq {\sf M}
\]
\end{lemma}
\begin{proof}
  The proof is very simple. Every partition is schedulable if we do
  not consider the other partitions. Therefore the whole system is
  schedulable, if the total number of required resources is less or
  equal to the available resources.
\end{proof}

Lemma \ref{lemma:sched:all} allows to exit Algorithm \ref{CHalgorithm}
on success at any time of its execution. If the condition in Lemma
\ref{lemma:sched:all} is not satisfied, we move the step, namely {\it partition merging}.

\subsection{Partition merging and minimizing resources}

The goal of the second step of our heuristic is to reduce the number
of required SMs for the input task set to be schedulable. Even if the
number of required SMs per partition is \newtext{minimal (i.e. if the number of SMs is reduced by, the system becomes non-schedulable)}, the schedulability
test of Lemma \ref{lemma:sched:all} is pessimistic when we consider
the whole platform, as the maximum schedulable workload per every
partition is probably not reached.

\begin{definition}\label{def:validmerge}
  Let $\mathcal{P}^1(\mathcal{T}^1, m^1)$ and
  $\mathcal{P}^2(\mathcal{T}^2, m^2)$, be two partitions.
  
  $\mathcal{P}^3(\mathcal{T}^3, m^3)$ is a valid merge of  $\mathcal{P}^1$ and $\mathcal{P}^2$ if :

  \begin{align}
    \mathcal{T}^3 &=   \mathcal{T}^1 \cup \mathcal{T}^2 \\
    m^3 &< m^1 + m^2
  \end{align}
  and the merge is denoted as : $\mathcal{P}^3 = \mathcal{P}^1 \bigoplus \mathcal{P}^2$
\end{definition}

According to Definition \ref{def:validmerge}, a merge is considered as
valid only if it allows to allocate all the tasks of two partitions to a 
single new partition and that the required resources to schedule the
merged task set, is less than the sum of the required ones for the two partitions.

\begin{algorithm}[h]
\caption{\sf merge}
\label{algo:mrg_algorithm}
\begin{algorithmic}[1]
  \State {\bf Input} {Partition $\mathcal{P}^1, \mathcal{P}^2$}
  \State schedulable = false;
  \State $\mathcal{T}(\mathcal{P}^1 \bigoplus  \mathcal{P}^2)= \mathcal{T}(\mathcal{P}^1) \cup \mathcal{T}(\mathcal{P}^2) $
  \State $m=\max\{|\mathcal{P}^1|, |\mathcal{P}^2| \} $
  \While {({\bf not} schedulable) {\bf and} $m < |P1| + |P2|$}
      \State $schedulable = test\_schedulability(\mathcal{T}^{tmp}, m); $
            \If {(schedulable)}
                \State $save\_new\_cluster(\mathcal{T},m)$
                \State $remove\_clusters(P1,P2);$
            \State  {\bf return} {\sf true}
          \Else
             \State $m=m+1;$
          \EndIf
  \EndWhile
  \State {\bf return } {\sf false}
\end{algorithmic}
\end{algorithm}

Algorithm \ref{algo:mrg_algorithm} decides whether a merge is valid or
not. It takes as input two partitions and produces a valid partition
or fails. It first merges the two task sets (Line 3). Further, it
tests iteratively the schedulability of the merged task sets. If the
schedulability fails, it increases the number of required resources
and retests the schedulability (Line 6). It considers in the first iteration a number of resources equal to the maximum number of resources between
the two input partitions (Line 4). Iteratively, the number
of required resources is increased (Line 12). This algorithm can be improved by using a binary
search, applied between $\max\{|\mathcal{P}|^1, |\mathcal{P}|^2 \}$
and $|\mathcal{P}^1| + |\mathcal{P}^2|$. The algorithm exits when the
number of required resources $m$ is greater or equal to the required
resources of the two partitions when considered independently.

A merge can fail because of the increase of the execution time when
two conflicting tasks are merged and therefore require more resources
than when the partitions are independent. It can also fail if it does
not allow to reduce the number of required resources (the merge leads
to a number of resource that is equal to sum of resources of the two
partitions), in this situation it is better to keep each partition
independent, as it contains less tasks, and therefore allows finer merges in future iterations.

\begin{definition}{(Order relation $\gg$)}\label{def:order:partial}

  Let $\mathcal{P},\mathcal{P}'$ and $\mathcal{P}''$  be three partitions.

  We define partial relation order $\gg$ as follows :
  
  \[
    \mathcal{P} \bigoplus \mathcal{P}' \gg \mathcal{P} \bigoplus \mathcal{P}'' \implies |\mathcal{P} \bigoplus \mathcal{P}'| < |\mathcal{P} \bigoplus \mathcal{P}''|
    \]

\end{definition}

The order relation $\gg$ allows to order two merges, where the same partition is involved, as a function of the resources they use.

\begin{definition}{(Order relation $>$)}\label{def:order:total}

  Let $\mathcal{P},\mathcal{P}'$ and $\mathcal{P}''$  be three partitions.

  We define partial relation order $>$ as follows :
  
  \[
    \mathcal{P} \bigoplus \mathcal{P}' > \mathcal{P} \bigoplus \mathcal{P}'' \implies U(|\mathcal{P} \bigoplus \mathcal{P}'|) < U(|\mathcal{P} \bigoplus \mathcal{P}''|)
    \]

Where $U$ is the partition utilization, defined as follows : 
\[
U(\mathcal{P}) = \sum_{i=1}^{|\mathcal{P}|} C_i(\mathcal{T}^p - \{\tau_i\}, |\mathcal{P}|)
\]
  \end{definition}

The order relation $>$ allows to order two merges, where the same partition is involved, as a function
of the workload that they generate.

Order relation $>$ and $\gg$ will be used in the experimental section to evaluate their performances of our heuristics.

\newtext{
Another aspect to consider is the \textit{tail effect}. Tail effect occurs when the last blocks of a kernel unevenly distributes among the available SMs. To understand this effect, we stress that 
when a grid of threads for a kernel is launched, that grid is divided into \textit{waves} of thread blocks. The size of a wave depends on the number of SMs on the GPU and the theoretical occupancy of the kernel~\cite{sched_hierarchy_RTAS}. Waves tend to be executed simultaneously within all the SMs. A sub-optimal configuration of number of blocks, therefore can cause the last wave to be served by a small number of SMs, hence this last wave causes to increase the execution time of the whole kernel while under-utilizing the GPU
\footnote{More details on tail effect and its implications can be found here \url{https://developer.nvidia.com/blog/cuda-pro-tip-minimize-the-tail-effect/}}.

}
The tail effect can also be an important factor when defining the
merge to select, and can condition the performance of the heuristics. If the number of blocks of the different kernels are multiples, it is worth
defining the partition size as a divider of the base period, to avoid tail effects.  However,
in this work we consider general case, where the number of blocks per
task is arbitrary. Therefore resizing the partition to eliminate the
tail effect for a given task may worsen it for other tasks. Hence,
it is not considered in this work.

\subsection{Forbidden list}

\newtext{The ${\sf forbidden\_list}$ is a list of task pairs. The tasks within the same pair \emph{usually} require more resources to meet deadlines, when allocated to the same cluster, than when each is allocated to a  different cluster. Therefore,} when a merge fails, the input partitions are added to the forbidden
list. The \newtext{${\sf forbidden\_list}$ is checked before selecting the partitions to merge within ${ \sf select\_partitions}$ (Algorithm \ref{algo:select_par__algorithm}).}

${\sf forbidden\_list}$ is first initialized as empty and is filled during analysis when a merge fails. An additional step,  denoted by ${\sf ACT}$ in Section \ref{sec:results},  and by ${\sf INA}$ when  not used, allow to fill it explicitly as a preliminary step.  In the later case, every couple of tasks is tested to check if they are mergeable. If the test fails, the couple of tasks is added to the ${\sf forbidden\_list}$. 


\subsection{Select partitions}
The partition selection process is very important to the performance of Algorithm \ref{CHalgorithm}, as our heuristic do not back-track the design space exploration.  The partition selection (Algorithm \ref{algo:select_par__algorithm}) goes through 3 steps. The first is to select the first operand partition $\mathcal{P}$ of the merge operation (Line 5). Further, it selects all candidate partitions, by excluding those where at least one task is implied in a forbidden merge with a task of  $\mathcal{P}$. ${\sf elig\_list}$ contains the list of eligible partitions to the merge with $\mathcal{P}$. This eligible list is first sorted according to either $\gg$ or $>$.

\begin{algorithm}
\caption{select\_partitions}
\label{algo:select_par__algorithm}
\begin{algorithmic}[1]
\State ${\bf Input~} {\sf list~of~partition : par\_list}$
\State ~~~~~~~~~~${\sf order} : \{ \gg or > \}$
\State ${\sf candidates} = {\sf par\_list}$ 
\Repeat
  \State $\mathcal{P} = {\sf choose\_from} ({\sf candidates})$
  \State ${\sf candidates} = {\sf candidates} \setminus \{ \mathcal{P}\}$
  \State ${\sf forbidden} = \{ \mathcal{P}' \mid (\mathcal{P}, \mathcal{P}') \in {\sf forbidden\_list} \lor (\mathcal{P}',\mathcal{P}) \in {\sf forbidden\_list}\}$
  \State ${\sf elig\_list} = {\sf par\_list} \setminus {\sf forbidden}$
  \If {${\sf elig\_list} \neq \varnothing$}
  \State {\sf return} $(\mathcal{P}, {\sf sort(elig\_list, order)})$
  \EndIf
 \Until ${\sf part\_list} = \varnothing$
 \State ${\sf return~NULL}$
\end{algorithmic}
\end{algorithm}

By definition, $\gg$ relation orders merged partitions, therefore it returns only one element containing the best merge. While $>$ order relation does not require a merge to sort partitions and therefore returns a list of sorted partition list (${elig\_list}$).

\subsection{Schedulability test}

The scheduling policy within each partition is independent of the proposed partition and task to partition allocation:
any policy for which there is an effective schedulability test can be used.

Choosing a specific scheduling policy allows us to further reduce the complexity of defining partition sizes. It also allows us to increase every partition utilization, as well as refining the selection of ${\sf elig\_list}$, by applying time sensitive analysis as those found in \cite{zahaf_ieee_tc_2019} \cite{zahaf_jsa_17}.

Independently from scheduling policy, we highlight two approaches for taming the interference effects of conflicting kernels within the same partition. 

The first one consists in explicitly taking into account for the interference in the analysis. 
This is a non trivial problem which is still a work in progress, thus, it will be detailed in a future work.

The second, more pessimistic but also simpler approach, consists in accounting for the effect of interference in the WCETs estimations. This is the approach we followed in this work. 
Let us recall that in the considered model, the function that gives the worst case execution time of a task takes into account the absence (resp. presence) of conflict for this task within its partition to choose a duration that does not integrate (resp. integrate) the effect of interference.


    




\section{Related work}\label{relatedWork}
When considering multi-kernel execution,
previous work highlights that improving GPU utilization can be achieved through two different mechanisms, namely \textit{Spatial Multikernel or Spatial Partitioning}  \cite{gpu_multitasking} and \textit{Simultaneous Multikernel (SMK)} \cite{SMK}.
In SMK, multiple kernels are scheduled in the same SMs, hence, kernels contend the compute resources. In this sense, in a time window, only one kernel can occupy the GPU computing resources. 
In terms of optimizing utilization, this approach is advisable when smaller GPUs are involved (i.e. low SM count), as in those devices a single kernel is very likely to occupy the entire GPU computing resources. In this work, however, we target current and next generation GPUs featuring a SM count such to make this approach sub-optimal.
For this reason, our contribution focuses on the second approach for concurrent kernels arbitration. In the approaches based on spatial partitioning, multiple kernels execute at the same time in different "SM-partitions" by sharing both compute and memory resources. 

\cite{gpu_multitasking} is one of the first works to make the case that spatial partitioning results in higher performance gains than time partitioning mechanisms. The authors of \cite{gpu_multitasking} classify workloads as compute-bound, memory/interconnect-bound and problem-size bound depending on their performance sensitivity to varying number of SMs, memory and interconnect frequency. They also evaluate various compile-time spatial partitioning schemes. 

The authors of \cite{classification_driven} present a scheme to dynamically assign SM partitions to kernels. This approach is iterative, meaning that a decreasing number of SMs is assigned to a kernel; specifically, authors aim to assign the smallest number of SM per kernel without degrading the overall performance.

Similarly, \cite{classification_driven2} also focuses on dynamic spatial partitioning of the GPU for cloud-based systems. Authors aim to achieve fairness and QoS targets. They propose a model to predict the slowdown caused by spatial multitasking to guide the dynamic partitioning scheme. The paper illustrates two partitioning schemes. The first scheme, HSM-Fair iteratively repartitions until the normalized progress of each running kernel is approximately equal. Whereas, HSM-QoS ensures that all high priority kernels meet their normalized progress thresholds and then maximizes for the overall system throughput.  

The authors of \cite{slate} propose a software based kernel scheduler, Slate. Slate finds complementary resource demands to co-schedule kernels while minimizing the interference between concurrent kernels.
Hence, they profile the performance of concurrent kernels to understand whether two kernels are to be considered complementary. Ideally, a compute bound and a memory bound kernels are considered complementary, as they won't interfere each other.
Unfortunately, in the Slate paper, they do not discuss how to spatially partition SMs for each of the co-running kernel.


In \cite{smqos}, Sun et al. present SMQoS a software based mechanism to dynamically adjust SM allocation. The algorithm proposed aims for maximizing throughput using information obtained by co-running tasks. 
In \cite{themis}, authors present a similar SM allocation engine that relies on the slowdown of the co-running applications to determine the number of SMs allocated to each of the kernels. Authors propose two different heuristics, one based on fairness and the other in quality of service.

None of the above mentioned papers consider real-time constraints as periods or deadlines in the partitioning problem. Moreover, many of the proposed methods in the literature just focus on pair-wise allocation methods where just two kernels are considered as concurrent elements. The GPU has been partitioned where each SM is a different partition in \cite{pruda2} \cite{pruda1}. However, the real-time software developer has to design its kernels, so their timing requirement are satisfied by a single SM.

\begin{figure*}[t]
  \centering
    \resizebox{0.45\textwidth}{!}{
  \begin{tikzpicture}
    \begin{axis}
      [
      axis x line = bottom,
      axis y line = left,
      xlabel={Total utilization},
      ylabel={Schedulability rate},
      xmax = 60,
      ymax = 1.1,
      xmin = 35,
      ymin = 0,
      legend entries={1G, SMS\_ACT,SMS\_INA, BF\_ACT, BF\_INA},
      y label style={at={(axis description cs:0.05,.5)},rotate=0,anchor=south},
      legend style={at={(1.15,1)}}
      ]
      \addplot [color=black,dashed, mark= diamond*] table[y index=2,x index=1]{texfiles/results/new/res50/1G_avg.dat};
      \addplot [color=red!90,dashed, mark=*,mark options={scale=1, fill=white} ]  table[y index=2,x index=1]{texfiles/results/new/res50/SMS_ACT_avg.dat};
      \addplot [color=red, mark= square*] table[y index=2,x index=1]{texfiles/results/new//res50/SMS_INA_avg.dat};
      \addplot [color=blue!90,dashed, mark=*] table[y index=2,x index=1]{texfiles/results/new//res50/BF_ACT_avg.dat};
      \addplot [color=blue, mark=+] table[y index=2,x index=1]{texfiles/results/new//res50/BF_INA_avg.dat};
    \end{axis}
  \end{tikzpicture} 
  }
  \quad
  \resizebox{0.45\textwidth}{!}{
   \begin{tikzpicture}
    \begin{axis}
      [
      axis x line = bottom,
      axis y line = left,
      xlabel={Total utilization},
      ylabel={Schedulability rate},
      xmax = 55,
      ymax = 1.1,
      xmin = 45,
      ymin = 0,
      legend entries={SMS\_ACT,SMS\_INA, BF\_ACT, BF\_INA},
      y label style={at={(axis description cs:0.05,.5)},rotate=0,anchor=south},
      legend style={at={(1.15,1)}}
      ]
      
      \addplot [color=red!90,dashed, mark=*,mark options={scale=1, fill=white} ]  table[y index=2,x index=1]{texfiles/results/new/res50/SMS_ACT_avg.dat};
      \addplot [color=red, mark= square*] table[y index=2,x index=1]{texfiles/results/new//res50/SMS_INA_avg.dat};
      \addplot [color=blue!90,dashed, mark=*] table[y index=2,x index=1]{texfiles/results/new//res50/BF_ACT_avg.dat};
      \addplot [color=blue, mark=+] table[y index=2,x index=1]{texfiles/results/new//res50/BF_INA_avg.dat};
      
    \end{axis}
  \end{tikzpicture}
  }
  
  \caption{(a) Schedulability rate for 50 tasks, with $prm = 50\%$} (b) zoom-in schedulability rate\label{fig:sched:50:50n}
\end{figure*}
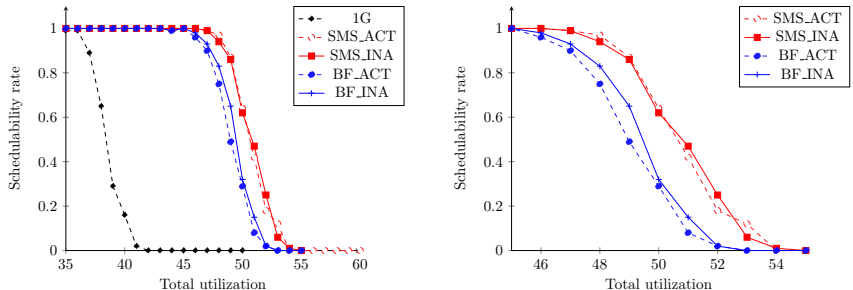

\section{Results and discussions}\label{results}
\label{sec:results}

In this section, we evaluate the performance of different versions of the proposed heuristics against using the GPU as a single non-partionable resource. Our simulations consider the NVIDIA GeFORCE 2080 Ti GPU, featuring $68$ compute resource (SMs), on a set of synthetic tasks.

\subsection{Task set generation}

We apply the proposed heuristics on a large number of randomly generated synthetic tasksets.

Initially, the number of tasks is fixed according to two scenarios: in the first scenario, we generate 50 tasks, i.e assigning an average load of 0.73 task per SM.
The goal in this first scenario is to select the number of tasks that is less than the number of SMs, so that it will enforce our heuristic to miss-behave. In fact, in such cases,
the grouping capabilities of our approach are reduced, as it will likely create a partition for every task. As mentioned in Section \ref{heuristics}, we expect that the maximum schedulable per-partition workload will not be reached. In the second scenario, we generate 200 tasks. Therefore in average, $2.94$ tasks per-SM are assigned.

For every scenario, we start by first generating tasks utilizations
using the UUniFast-Discard algorithm \cite{techniques}.  The baseline
periods of every task are selected from a predefined list of periods,
where the lowest period is $50$, and the highest period is
$4000$. Such numbers have been chosen to allow us to study tractable hyperperiods.
When selecting the task period, we make sure that a reasonable execution time can be assigned to the task.
If such execution time is not reasonable, then the period is increased within the limit of $4000$. The task deadline is set to $0.75 \mathcal{T}_i$.

By multiplying the task period and the task utilization, we obtain the baseline execution time of each task.
Then, we use the following execution time scaling function: $\mathcal{C}_i(\mathcal{T}(\mathcal{P}),|\mathcal{P}|) = k_i \left( \frac{a_i}{|\mathcal{P}|} + b_i \right)$ where:
\begin{itemize}
    \item $a_i$ is the baseline execution time of $\tau_i$; 
    \item $b_i$ represents the part of the execution time that does not profit from parallelism. It is set to $0.02 \times a_i$ (resp. $0.1 \times a_i$) if $\tau_i$ is compute intensive (resp. memory intensive);
    \item $k_i$ is a factor used to inflate execution times in case of conflict. Thus it is set to $1$ when $\tau_i$ has no conflict in $\mathcal{P}$, and to $1.2$ (resp. $2.3$) when $\tau_i$ has conflict in $\mathcal{P}$ and is compute intensive (resp. memory intensive).
\end{itemize}

A task is chosen to be compute or memory intensive kernel according to the
$prm$ variable. A random number $\alpha$ is generated between $0$ and
$1$. If $\alpha$ is greater than $prm$, the task is a compute
intensive kernel, otherwise it is a memory intensive kernel. In the
different experiments, $prm$ is set to $ 50\%$ 

\subsection{Simulations}

We vary the total utilization from $2$ to $68$, with step of $2$. Each point
in the figures represents the average value among $100$ simulations.

We use three different metrics to evaluate the performance of our heuristics. The first is the classical schedulability rate, so that we count the number of schedulable tasksets among $100$ executions. The second metric is the scheduled workload. According to
the heuristic we use, we may select non-conflicting or conflicting execution time. The more non-conflicting execution time is selected, the better the heuristic is. The third metric is the number of partitions. The baseline
heuristic where the GPU is considered as single non-partionable resource is denoted in the Figure by ${\sf 1G}$. Our heuristics can
either use $\gg$ heuristics, which is denoted by ${\sf SMS}$ in the
figures, or use $>$ which represents the best fit algorithm, denoted
by ${\sf BF}$. The preliminary step where all possible pairs are
tested for a merge to fill an exhaustive forbidden pair list is denoted by
${\sf ACT}$. In contrast ${\sf INA}$ denotes heuristic that do not use this step (Line 3, Algorithm \ref{CHalgorithm}).

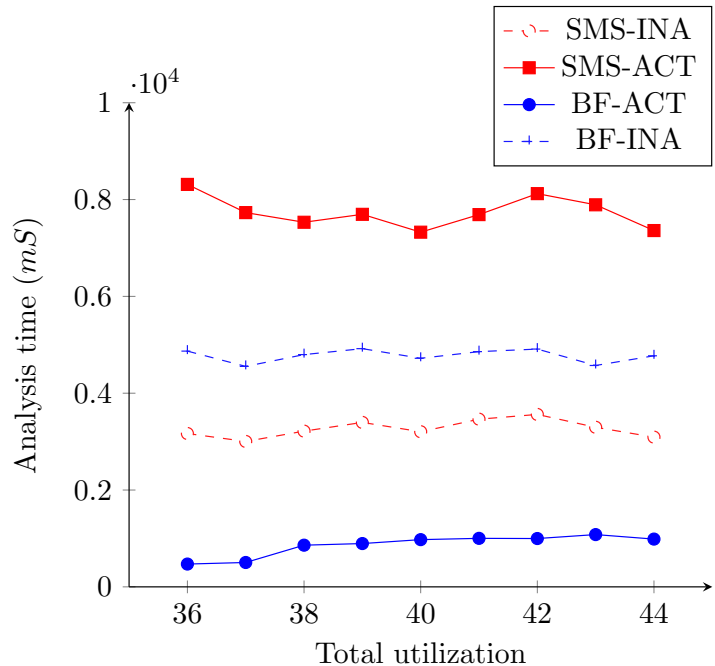
\begin{figure}[h]
  \centering
  \resizebox{0.8\columnwidth}{!}{
  \begin{tikzpicture}
    \begin{axis}
      [
      axis x line = bottom,
      axis y line = left,
      xlabel={Total utilization},
      ylabel={Analysis time ($mS$)},
     xmax = 45,
      ymax = 10000,
     xmin = 35,
     ymin = 0,
      legend entries={SMS-INA, SMS-ACT, BF-ACT, BF-INA},
      y label style={at={(axis description cs:0.05,.5)},rotate=0,anchor=south},
      legend style={at={(1,1.2)}}
      ]
      \addplot  [color=red!90,dashed, mark=*,mark options={scale=1, fill=white} ]  table[y index=2,x index=0]{texfiles/results/exec.dat};
       \addplot [color=red, mark= square*] table[y index=1,x index=0]{texfiles/results/exec.dat};

       \addplot [color=blue, mark=*]table[y index=3,x index=0]{texfiles/results/exec.dat};
       \addplot  [color=blue!90,dashed, mark=+]  table[y index=4,x index=0]{texfiles/results/exec.dat};
    
    \end{axis}
  \end{tikzpicture}
  }
  \caption{Analysis time as a function of total utilization}\label{fig:analysis:time}
\end{figure}

 In Figure \ref{fig:sched:50:50n}, we report the average schedulability as a function of total utilization. In this experiment, we have synthesized 100 tasksets per utilization factor, each having $50$ tasks, where each task has $50\%$ chance to be compute or memory intensive. When the total utilization is less than $35$, the GPU does not reach a critical load situation, and therefore all heuristics are able to achieve 100\% schedulability. Such a schedulability ratio is also reached by the baseline approach ($1G$), in which the whole GPU is considered as one single partition. Starting from $35$, the schedulability of $1G$ drastically falls, as it executes workloads close to maximal schedulable utilization. Indeed, when all tasks are in a single partition, they are probably in conflict, hence, their execution time with maximum interference is considered. In the right sub-figure, we zoom-in the schedulability of the proposed approaches in this paper. In fact, as the utilization increases, the $\gg$-based heuristics perform better than best fit heuristics. When the exhaustive forbidden pair list is activated, the different heuristics performance slightly downgrades. Indeed, the selection of candidates will have a limited number of  merge choices at early iterations of the algorithm. When it is disabled, the heuristics are allowed to merge tasks, that are not mergeable individually. These mergings are possible when the individual tasks belong to mergeable larger partitions. In other words, it allows the heuristics to build partitions including forbidden pairs through valid merging of larger partitions. This feature, which is implicitly prohibited when exhaustive forbidden pair list is used,  may lead to better solutions, as it  browses larger choices in the design space. However, when exhaustive forbidden pair list is activated, the execution time is considerably reduced (Figure \ref{fig:analysis:time}). 

\begin{figure}[h]
  \centering
    \resizebox{0.8\columnwidth}{!}{

  \begin{tikzpicture}
    \begin{axis}
      [
      axis x line = bottom,
      axis y line = left,
      xlabel={Total utilization},
      ylabel={Efficiency},
      xmax = 45,
      xmin = 35,
      ymin = 0,
      legend entries={Worst, Best, SMS\_ACT, BF\_ACT},
      y label style={at={(axis description cs:0.05,.5)},rotate=0,anchor=south},
      legend style={at={(1,0.4)}}
      ]
      \addplot  
      [color=gray,dashed, mark= square*] table[y index=3,x index=1]{texfiles/results/new/res50/SMS_ACT_avg.dat};
      \addplot [color=gray,dashed, mark= square*]  table[y index=4,x index=1]{texfiles/results/new/res50/SMS_ACT_avg.dat};
      \addplot [color=red!90,dashed, mark=*,mark options={scale=1, fill=white} ]  table[y index=5,x index=1]{texfiles/results/new/res50/SMS_ACT_avg.dat};
      \addplot [color=blue,dashed, mark= square*] table[y index=5,x index=1]{texfiles/results/new/res50/BF_ACT_avg.dat};
    \end{axis}
  \end{tikzpicture}
  }
  \caption{Clusturing efficiency rate for 200 tasks, with $prm = 50\%$}\label{fig:eff:50:50}
\end{figure}

In Figure \ref{fig:eff:50:50}, we report the efficiency of our heuristics in terms of partitioning the GPU for the same test benches as those of Figure \ref{fig:sched:50:50n}. 
The efficiency is computed as the ability of a heuristic to keep the utilization as low as possible by avoiding to build partitions with conflicting tasks.
Therefore, we first plot the efficiency bounds, which are computed for each taskset as its total utilization by using the non-interference execution times for the lower bound, and the interference execution times for the upper bound. 
The 1G heuristic always schedules the worst load as all tasks are within the same partition. 
Most of the time, our heuristics are able to find a solution such that each partition contains a single memory intensive and a single compute intensive tasks, therefore reducing the scheduled workload to minimum values. 
This is confirmed by the average number of partitions which is around $25$, hence, presenting $2$ tasks per partition. 
As BF efficiency is higher compared to SMS, it schedules more load than SMS, which corresponds to our previous discussion for Figure \ref{fig:sched:50:50n}.

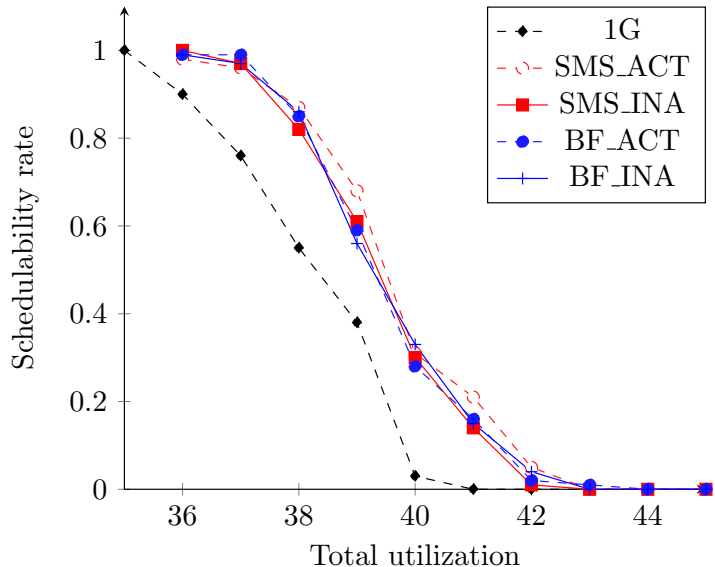
\begin{figure}[h]
  \centering
    \resizebox{0.8\columnwidth}{!}{
  \begin{tikzpicture}
    \begin{axis}
      [
      axis x line = bottom,
      axis y line = left,
      xlabel={Total utilization},
      ylabel={Schedulability rate},
      xmax = 45,
      ymax = 1.1,
      xmin = 35,
      ymin = 0,
      legend entries={1G, SMS\_ACT,SMS\_INA, BF\_ACT, BF\_INA},
      y label style={at={(axis description cs:0.05,.5)},rotate=0,anchor=south},
      legend style={at={(1,1)}}
      ]
      \addplot [color=black,dashed, mark= diamond*] table[y index=2,x index=1]{texfiles/results/new/1G_avg.dat};
      \addplot [color=red!90,dashed, mark=*,mark options={scale=1, fill=white} ]  table[y index=2,x index=1]{texfiles/results/new/SMS_ACT_avg.dat};
      \addplot [color=red, mark= square*] table[y index=2,x index=1]{texfiles/results/new/SMS_INA_avg.dat};
      \addplot [color=blue!90,dashed, mark=*] table[y index=2,x index=1]{texfiles/results/new/BF_ACT_avg.dat};
      \addplot [color=blue, mark=+] table[y index=2,x index=1]{texfiles/results/new/BF_INA_avg.dat};

    \end{axis}
  \end{tikzpicture}
  }
  \caption{Schedulability rate for 200 tasks, with $prm = 50\%$}\label{fig:sched:200:50}
\end{figure}

In Figure \ref{fig:sched:200:50}, we report the average schedulability as a function of total utilization for the similar experiments to those of Figure \ref{fig:sched:50:50n}, but by generating 200 tasks per task set rather than 50.

Again, our heuristics are dominating the non-partitioned GPU approach. We also note that for these experiments, our heuristics show performances closer to each other than in the previous results' set.  
Indeed, as tasks are smaller and more numerous, our heuristics are forced to allocate more than two tasks to each partition, which leads to increasing interference and thus downgrading their performances compared to the first experiments. 

\begin{figure}[h]
  \centering
  \resizebox{0.8\columnwidth}{!}{
  \begin{tikzpicture}
    \begin{axis}
      [
      axis x line = bottom,
      axis y line = left,
      xlabel={Total utilization},
      ylabel={Efficiency},
      xmax = 45,
      xmin = 35,
      ymin = 0,
      legend entries={Worst, Best, SMS\_ACT},
      y label style={at={(axis description cs:0.05,.5)},rotate=0,anchor=south},
      legend style={at={(1,0.4)}}
      ]

      \addplot [color=gray,dashed, mark= square*] table[y index=3,x index=1]{texfiles/results/new/SMS_ACT_avg.dat};
      \addplot [color=gray,dashed, mark= square*] table[y index=4,x index=1]{texfiles/results/new/SMS_ACT_avg.dat};
      \addplot [color=red!90,dashed, mark=*,mark options={scale=1, fill=white} ]  table[y index=5,x index=1]{texfiles/results/new/SMS_ACT_avg.dat};

    \end{axis}
  \end{tikzpicture}
  }
  \caption{Clustering efficiency for 200 tasks, with $prm = 50\%$}\label{fig:eff:200:50}
\end{figure}
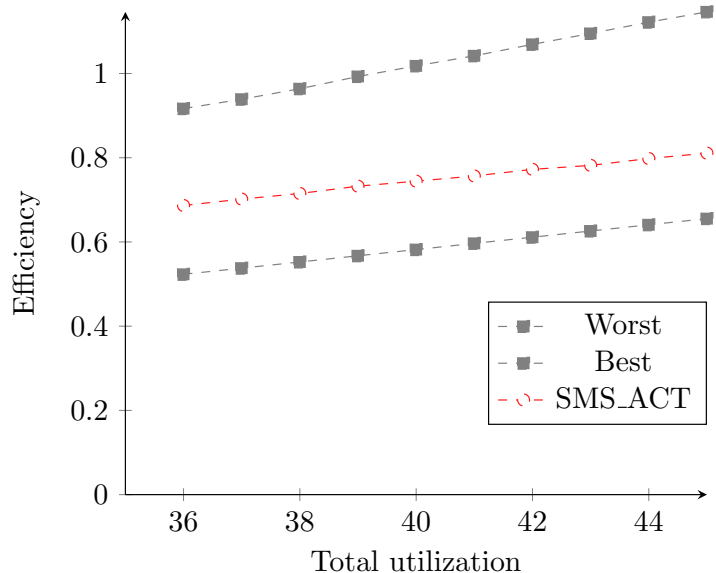
\noindent This is also demonstrated by Figure \ref{fig:eff:200:50} where the efficiency of our heuristics is worse compared to the previous experiments. Indeed, with more than 4 tasks allocated in average per partition, our heuristics are enforced to allocate conflicting tasks together. However, they still show a very large improvement regarding the $1G$ heuristic that always schedules the worst possible load.

In Figure \ref{fig:analysis:time}, we report the analysis time as a function of total utilization. 
It is notable that the analysis time does not depend on the total utilization. 
As expected, using the exhaustive forbidden pair list reduces the analysis time significantly. 
We can also notice that the heuristics based on best fit are faster than those based on SMS.
This is because $\gg$ relation used in SMS requires to compute candidate merging, while $>$ does not.

\section{Conclusion}\label{conc}

In this paper we proposed a GPU partitioning and
task-to-partition allocation mechanism able to account for kernels' timing constraints.  
Our heuristic takes into account the interference experienced from concurrent kernels as they compete on the same resources. We
demonstrate the performances of our heuristics against different
version of partitioning schemes and task-to-partition allocation mechanisms. We also compared our proposed heuristics against the baseline mechanism in which the GPU is considered  
as a single, non-partionable resource.

\newtext{
Our presented partitioning mechanisms is a research effort rather than only being an engineering effort. We therefore investigate approaches able to deal with a generalization of the problem within a fixed size; the mechanisms we detailed in this paper are able to be exploited by existing and future GPU programming models and frameworks with minimal modifications. For instance, our mechanisms can be applied on top of the NVIDIA MiG approach. In this case, the output of our method should be constrained within the admissible partitions made available with MiG. This is because MiG documentation states that partitions will be limited to specific configurations of SM/GPC groups, whereas our method considers partitions of arbitrary size.
Moreover, the memory partitioning features made available by MiG would imply tuning our model to consider every kernel as compute intensive. Other examples of approaches in which our proposed method can be used is the AMD's CU masking feature~\cite{otterness2020amd}.  
}

We plan to further extend this work by considering the interference and
the scheduling constraints that GPU kernels might experience when other are 
tasks running within the different processors/accelerators in the same computing platform. 
The goal is to integrate this work within the scheduling framework for heterogeneous
architectures developed in \cite{tc2020}. We also plan to define more
accurate kernels' categorization by accounting for architectural metrics such as cache-hit/miss rates, as profiled for each kernel.

\section{Akownledgment}
This research is a joint collaboration between researchers that have received funding from the SPHERE project (PRIN 2017, CUP: E94I19000820001) and the European Union’s ECSEL JU NEW CONTROL Project (agreement n. 826653).


\bibliographystyle{plain}
\bibliography{bibliography}
\end{document}